# The Nature of Schrödinger Equation
## On Quantum Physics Part I

Xue-Shu Zhao, Yu-Ru Ge, Xin Zhao and Hong Zhao

**Abstract.** We propose that the Schrödinger equation results from applying the classical wave equation to describe a physical system in which subatomic particles play random motion, thereby leading to quantum mechanics. The physical reality described by the wave function is subatomic particle moving randomly. Therefore, the characteristics of quantum mechanics have a dual nature, one of them is the deterministic nature carried over from classical physics, and the other is the probabilistic nature coined by particle's random motion. Based on this model, almost all of open questions in quantum mechanics can be explained consistently, which include the particle-wave duality, the principle of quantum superposition, interference pattern of double-slit experiments, and the boundary between classical world and quantum world. The current quantum mechanics is a mixture of matrix mechanics and wave mechanics, which are sharply conflicting in principle. Matrix mechanics treats quantum particles as classical particles with fixed relation between the particle's position and its momentum. The matrix mechanics, in fact, belongs to the old quantum theory. Both Born's non-commutative relation and Heisenberg uncertainty relation originate from matrix mechanics. However, in wave mechanics, there is no any fixed relation between the particle's position and its momentum, and the particle's position and its momentum belong to immeasurable physical quantities. Therefore, there is no need for non-commutative relation and uncertainty relation in wave mechanics.

## I  Introduction

Throughout human history, Newtonian laws of nature have provided people with a consistent and coherent picture of the nature world. It is believed that all science consists of an endless sequence of causes and effects. If causality no longer works, then true science ceases to exist.

In the early $20^{th}$ century, two articles by Einstein ushered in the era of modern physics. One of them is his special theory of relativity, which applies to particles traveling at close to the speed of light. The other is the theory of light quanta based on the theory of Max Plank, a pioneer of quantum world. Einstein first adopted the view that light has dual nature when explaining the photoelectric effect. In other words, light can act like an electromagnetic wave, so its energy can be distributed over a large area of space, but light can also act like quanta (photon) whose total energy must be confined within a very tiny volume. Placing two contradictory concepts on the same reality is bound to shake the foundation of science.

In 1923, Louis de Broglie postulated that perhaps all forms of matter possess wave and as well as particle properties. This postulate makes the situation even more counterintuitive. After Heisenberg's uncertainty principle was accepted by quantum world as the core of quantum mechanics, all theories and concepts based on classical physics were demolished. The laws of conservation of energy and momentum were abandoned. It is believed that the laws for causality and determinism no longer work; it should not be applied even in classical physics, the determinism law is meaningless and can only talk about probabilities of outcomes [1].

Copenhagen interpretation believes that in the measurement of a certain physical quantity, it is impossible to exclude people and the measuring equipment, which means that scientific objectivity no longer exists [2]. Obviously, the foundation of modern physics is completely separated from classical physics. Einstein was also shocked: "all my attempts to adapt the theoretical foundation of physics to this knowledge have failed completely. It is as if the ground has been pulled out from under one, with no firm foundation to be seen anywhere upon which one could built" [3].

Max Born, one of the founders of quantum theory, said "We are in the jungle; we find our way through trial and error, building our road behind us as we proceed" [4]. A century later, we still seem to be stuck in the jungle, and struggling to find the right way out of the jungle. However, the situation we face is much worse than our founders faced, since the 'GPS 'of classical physics has been shut down by our founders. Fortunately, we have the Schrödinger equation, a brilliant bridge built on the foundation of classical physics. Along it, we have the opportunity to bridge the chasm between the classical and quantum worlds.

Now, we must ask ourselves a fundamental question: what is the exact knowledge on which the quantum world is based? Does the quantum world have its own foundation? We must rigorously reconsider the theories, concepts and models that our funders have left us. We need to ensure that they are compatible and unified with the fundamental laws of physics.

In this article, we demonstrate that the essential nature of quantum particles is their random movement, which coins the probabilistic nature on quantum mechanics. Thus, the reality described by the wave function of the Schrödinger equation is a physical system in which particles move randomly. A reasonable way to get the Schrödinger equation lies in applying the deterministic classical wave equation to describe the system of random motion of subatomic particles. Therefore, the characteristics of quantum mechanics have dual nature; On the one hand, quantum mechanics inherits the determinism of classical physics, which is why quantum mechanics' predictions of state energies can be experimentally verified with extremely high accuracy. On the other hand, quantum mechanics has probabilistic nature caused by random motions of particles.

Based on this new model, a number of unsolved questions in quantum mechanics, such as particle-waves duality, the principle of quantum superposition, entanglement effects, double-slit interference pattern, and the boundary between quantum and classical worlds, can be explained consistently. Even so, we still cannot completely explain the R. Feynman's famous quote "Nobody understands quantum mechanics" [5]. The reason is that the quantum mechanics we currently accept is a mixture of matrix mechanics and wave mechanics, and the two mechanics are sharply conflicting in principle. The matrix mechanics treats quantum particles as classical particles with a fixed relation between a particle's position and its momentum. The position and momentum of particles belong to measurable physical quantities that dominate the particle's energy state.

In wave mechanics, however, the reality described by wave function is the random motion of quantum particles. The position and momentum of particles must be independent of each other and belong to immeasurable physical quantities. Because in wave mechanics, we only know the probability of a particle appearing at a given location, it is meaningless to measure accurately these two physical quantities. Thus, there is no need for Heisenberg uncertainty relations in quantum mechanics.

A fundamental mistake made in quantum mechanics is to regard the independent coordinate and time variables of the wave function as quantum operators, which leads to two non-commuting relations $[\hat{p}, \hat{x}]$ = -i$\hbar$ and $[\hat{E}, \hat{t}]$ = -i$\hbar$. If neither of the two non-commuting relations holds, then Heisenberg uncertainty relations have no a foothold in quantum mechanics.

In the discussion section at the end of this paper, we propose that in both classical and quantum systems, the quantization of energy can only occur in physical system where particles move randomly. Moreover, the root of energy quantization lies in the discontinuous changes of microstates in both classical and quantum systems.

**II  Schrödinger Equation and the Principle of Quantum Superposition**

It is well known that temperature field exists everywhere in the universe even in deep space far from Galaxies. The tiny particles in a temperature field will move randomly in all directions with equal probability and various speeds. The average kinetic energy of a particle of mass m in an environment of temperature T is

$$\frac{1}{2} m\overline{v^2} = \frac{3}{2} kT \qquad (1)$$

Where k is the Boltzmann constant, the square root of $\overline{v^2}$, $(\overline{v^2})^{1/2}$, is called the root mean square (rms) speed of the particle, which is close to the most probable speed at the same temperature. From (1) we get

$$v_{rms} = (\overline{v^2})^{1/2} = \left(\frac{3kT}{m}\right)^{1/2} \qquad (2)$$

Here $v_{rms}$ is the rms speed of a particle with mass m. The equation (2) shows that the rms speed of a particle is proportional to the square root of temperature and inversely proportional to the square root of mass.

At a given temperature, the speed ($v_{rms}$) of lighter particles is faster than that of heavier particles. Table 1 lists the rms speeds of some subatomic particles, and their corresponding de Broglie wavelengths at 20°C. It can be seen from the Table 1 that the root mean square speed of electrons at room temperature is 1.15 x $10^5$ m/s, approaching one thousandth of the speed of light. At a given temperature, in the absence of any disturbing field, an electron will move randomly around its initial position $x_0$ with a root mean square speed (Figure.1). Since movement in all directions has the same probability, the trajectory distribution pattern generated by an electron moving randomly should have spherical symmetry about its initial position $x_0$.

Table 1, the $\upsilon_{rms}$ speed and L. de Broglie wavelength $\lambda_{db}$

| Particle | mass (amu) | $\upsilon_{rms}$ at 20°C (m/s) | $\upsilon_{rms}$ at 1000°C (m/s) | $\lambda_{db}$ (m) |
|---|---|---|---|---|
| Electron | $5.5 \times 10^{-4}$ | $1.15 \times 10^5$ | $2.4 \times 10^5$ | $6.6 \times 10^{-9}$ |
| Proton | 1 | $2.7 \times 10^3$ | $5.6 \times 10^3$ | $1.5 \times 10^{-10}$ |
| $N_2$ | 28 | $5.1 \times 10^2$ | $1.07 \times 10^3$ | $2.8 \times 10^{-11}$ |
| $C_{60}$ | 720 | 100 | 209 | $5.5 \times 10^{-12}$ |
| Si particle with $5 \times 10^3$ atoms | $1.4 \times 10^5$ | 7.2 | 15 | $3.9 \times 10^{-13}$ |
| Si particle with $10^4$ atoms | $2.8 \times 10^5$ | 5.1 | 10.7 | $2.8 \times 10^{-13}$ |

To simplify the situation, Figure 1 shows a schematic diagram of the instantaneous trajectory distribution of the random movement of an electron confined in x – y plane around its initial position $x_0$.

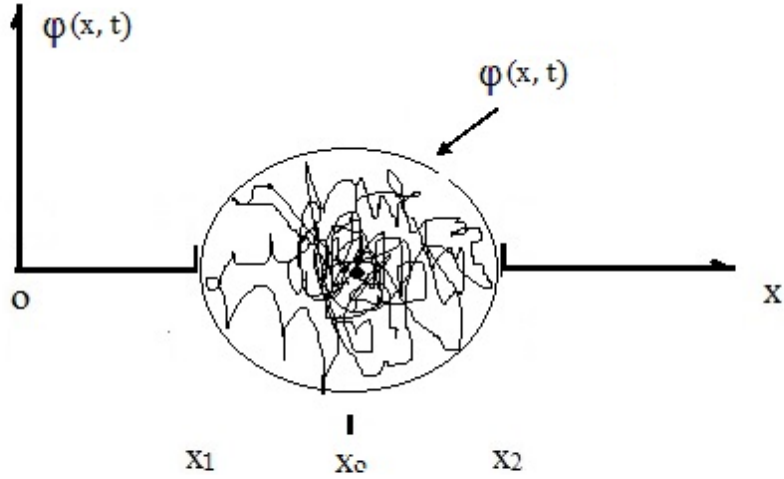

Fig.1. The schematic diagram shows that at a particular time, an electron confined in the x-y plane moves randomly outward from its initial position $x_0$ over time.

Now we use $\varphi(x, t)$ to represent the one-dimension electron distribution function describing the electron moving randomly within its wave packet. At a particular time, when the coordinate is beyond the interval $x_1$ to $x_2$, the distribution function $\varphi(x, t)$ should equal to zero. In this situation, the electron distribution function $\varphi(x, t)$ looks like standing wave fixed at the both ends of the string. The important feature of the standing wave function is that the complete wave function $\varphi(x, t)$ can be written as the product of two terms, one of them depends only on $x$, and another depends only on t. So that the complete electron wave function has the form

$$\varphi(x, t) = \varphi(x)e^{-i\omega t} \qquad (3)$$

In classical physics, for waves travelling along $x$ axis, the general form of the wave equation is

$$\frac{\partial^2 \varphi(x,t)}{\partial x^2} = \frac{1}{v^2}\frac{\partial^2 \varphi(x,t)}{\partial t^2} \qquad (4)$$

Where $v$ is the wave speed, if we substitute above wave function (3) into equation (4), we get

$$\frac{\partial^2 \varphi(x)}{\partial x^2} e^{-i\omega t} = -\left(\frac{\omega^2}{v^2}\right)\varphi(x)e^{-i\omega t}$$

Dividing both sides by $e^{-i\omega t}$, we get

$$\frac{\partial^2 \varphi(x)}{\partial x^2} = -\left(\frac{\omega}{v}\right)^2 \varphi(x) \qquad (5)$$

The equation (5) only contains variable $x$. Recall Einstein hypothesis that the energy and momentum of a photon are $E = \hbar\omega$ and $p = h/\lambda$, respectively. Since the energy and momentum of a photon are originated from the transition of an electron from its higher energy level to a lower energy level, the transition process must satisfy both the law of conservation of energy and the law of conservation of momentum at the same time. Therefore, it is reasonable to apply the equations of the photon energy and of its momentum to the corresponding electron physical quantities. Using formula of $\omega = 2\pi v/\lambda$ and $p = h/\lambda$ we get

$$\frac{\omega^2}{v^2} = \frac{p^2}{\hbar^2}$$

Where ω is angular frequency, p is momentum of electrons, $\hbar$ is the reduced Plank's constant. In classical physics, the total energy E of a system is the sum of the kinetic energy $\frac{p^2}{2m}$, and the potential energy U, which can be expressed by

$$E = \frac{p^2}{2m} + U \quad \text{or} \quad p^2 = 2m(E - U)$$

Therefore, we get

$$\frac{\omega^2}{v^2} = \frac{p^2}{\hbar^2} = \frac{2m}{\hbar^2}(E-U)$$

Substituting above equation into equation (5) gives

$$\frac{\hbar^2}{2m}\frac{\partial^2 \varphi(x)}{\partial x^2} + (E - U)\varphi(x) = 0 \tag{6}$$

The equation (6) is the famous one –dimension time- independent Schrödinger equation for electrons, which has enormous successes for precisely determining the structure of atoms, molecular, and solids.

From the path for deriving time-independent Schrödinger equation, we can see clearly that the wave function of Schrödinger equation is no longer explained like Copenhagen's interpretation; it is just an abstract mathematical structure, having nothing to do with the reality of the described system. The above derivation process gives a straightforward connection between classical and quantum worlds. The physical reality described by wave function $\varphi(x)$ is the randomly moving electron hidden in its wave function (shown as Figure 1).

Now we can come to conclusions naturally that the Schrödinger equation is rooted in the solid foundation of classical physics. This equation is the natural result of applying deterministic classical mechanics to describe the system in which subatomic particle moves randomly. This is why the Schrödinger equation can provide extremely precise results for extensive experiments.

In addition, Equation (6) shows that the electron's energy E seems to be a free parameter, inserting any value of E into the equation and solving for φ(x) seems acceptable. However, to make it physically acceptable, the wave function must meet the normalization requirement resulting from the probability interpretation of the wave function. As suggested by Born, one of the founders of quantum mechanics, the square of the absolute value of the wave function φ(x) is defined as the probability of finding an electron at position x [1]. The normalization requirement means that the integral of the square of the absolute value of the wave function must be equal to 1 in the region required by the boundary conditions. The physically acceptable wave function must satisfy the normalization condition. Therefore, the particle will never have any energy other than those particular values allowed by the acceptable wave function; this characteristic is called energy quantization in subatomic system. Now we have obtained the significant result that the energy quantization of subatomic particle is dominated by the probability nature of the wave function, while the probability interpretation roots in the random motions of particles. Therefore, the energy quantization in quantum mechanics must originate from the random motions of subatomic particles. In other words, the exact knowledge on which the quantum world is based is the nature of quantum particles playing random movement.

The difference from classical mechanics is that in quantum mechanics, due to the particle's random motion, there is no longer a causal relationship between the electron's position and its momentum. In other words, even if the particle's position and its momentum are known at now, its past movement trajectory cannot be found, and its moving trajectory in future cannot be certainly predicted. This is the just reason based on which Copenhagen team believed that the laws of causality and determinism must be abandoned in the entire physics [1]. However, there is a rigorous causal relationship between wave function φ(x) and the system energy E in quantum mechanics.

According to the data listed in Table 1 above, it is obvious that subatomic particles are always in a random motion state in any environment except for absolute zero temperature, which can be fully described by the Schrödinger's equation. Since the Schrödinger equation can only deal with the random movement of subatomic particles, it does not provide any defined values for other physical quantities except for the energy of states, for those physical quantities, it can only give a prediction by using probability distribution; that is, for which it has to play dice.

If we take the time derivative of equation (3), we give

$$\frac{\partial \varphi(x,t)}{\partial t} = -i\omega \varphi(x, t),$$

Recall $E = \hbar\omega$, or $\omega = \frac{E}{\hbar}$

So
$$i\hbar \frac{\partial \varphi(x,t)}{\partial t} = E\, \varphi(x, t) \tag{7}$$

From equation (6), we have the total energy

$$E = -\frac{\hbar^2}{2m}\frac{\partial^2}{\partial x^2} + U \tag{8}$$

Substitute the equation (8) into (7), we get the famous time–dependent one- dimension Schrödinger equation:

$$i\hbar \frac{\partial \varphi(x,t)}{\partial t} = -\frac{\hbar^2}{2m}\frac{\partial^2 \varphi(x.t)}{\partial x^2} + U\, \varphi(x.t) \tag{9}$$

Generally, the solution of the time dependent Schrödinger equation (9) describes the dynamical behavior of quantum particles. On the other hand, Schrodinger equation is a linear equation, which means that it must satisfy the principle of superposition: that is, if functions φ$_1$ and φ$_2$ both represent valid solutions of the equation, their combination must also be valid solution. Thus, any linear combination of the solutions of the linear equation will also be its valid solution. If there are many arrangement states of particles in a physical system, then the most general solution should be a combination of all these possibilities. Obviously, the linear characteristic of Schrodinger equation is particularly useful for describing the physical system in which particles play random movement. The superposition principle of quantum mechanics has no analogs in classical physics. Most of equations in classical physics are non-linear equations;

they all have unique solution, which is associated with the deterministic nature of classical physics. Obviously, it is impossible in physics to derive strictly the Schrodinger equation from any principle of classical physics. That is, we cannot derive the linear Schrodinger equation from nonlinear classical equations. Although there are a number of ways to obtain this equation, none of them can provide a satisfying explanation for the linear feature of the Schrodinger equation.

In classical mechanics, the state of a moving particle can be objectively described by its position and momentum. If its initial condition is given, then moving trajectory of the particle will be completely fixed at all times. However, in quantum mechanics, the energy state of a subatomic particle is described by its wave function, in which the particle moves randomly. Particles can occupy any positions in the region constrained by the boundary condition, and can take any available momentum allowed by initial conditions. Under this situation, for a given wave function, there are unlimited numbers of moving trajectories of the particle, and each trajectory should have a wave function associated with it. These unlimited numbers of wave functions all belong to the same energy state. As a result, the linear combinations of all these unlimited numbers of wave functions can be regarded as a complete wave function belonging to the same energy state. This is just the physical origin of the famous quantum superposition principle in quantum mechanics, for which physicists has been looking for almost hundred years.

The analytical path to obtain the principle of quantum superposition naturally leads to the conclusion that the quantum superposition principle, the heart of quantum mechanics, also results from the same causality: subatomic particles are always in random movement state in any environment.

Coincidentally, the formula of the time-dependent Schrödinger equation (equation 9) is similar to the diffusion equation. If the diffusion coefficient is constant, then the diffusion equation is

$$\frac{\partial \phi(r,t)}{\partial t} = D \nabla^2 \phi(r, t) \qquad (10)$$

Where, $\phi(r, t)$ is the density of the diffusing material at location r and at time t. We know that the theories in physics must be consistent. The two equations (9) and (10) are linear differential equations and both describe the random motion of particles. The common nature of the two equations is to describe the physical process in which the energetic particles diffuse outward through random walk with the evolution of time. We can imagine that an energetic particle suddenly placed at position $x_o$, as shown in Figure 1, the particle will spread out in all directions, just as the wave function spreads out over time. The speed at which the wave function spreads depends on the particle's mass and its temperature. We will discuss this feature further in the double- slit interference experiments section.

We need to keep in mind that in classical physics, it is impossible to simulate the superposition of quantum mechanics in any form of waves; including mechanical waves, electromagnetic and light waves, they all have the deterministic nature. If two light waves or any other kind of waves are superimposed together, a new wave will be produced, the energy state of which is different from the two. Clearly, light acting as waves does not follow the superposition principle of quantum mechanics. In addition, light acting like a particle does not exhibit random motion, so that, the Schrödinger equation cannot be used to describe light particles. Thus, light acting as particles do not belong to quantum particles satisfying Schrodinger's equation. However, light particles do show interference pattern in doubt-slit experiment, which has been well explained by classical superposition principle.

## III   The Boundary between the Classical World and the Quantum World

For almost a century, physicists have struggled to find a clear boundary between classical and quantum world. Theoretically, it seems impossible to deal with this issue in the right path, because since founders began to constitute the quantum mechanics, the fundamental causality to realize the quantum mechanics has been ignored. The existing majority view is that since the quantum mechanics has never collapsed in large-scale world, theoretically, this boundary may not exist, and the laws of classical mechanics are an extension of the quantum laws in large-scale system [6]. However, experimentally, it has long been known that quantum particles can be in a superposition state and exhibit the interference pattern in the double-slit experiment. But does this also hold true for larger objects? If the interference pattern disappears at a certain large particle size in the double-slit experiments, then there must exist a boundary between classical and quantum worlds.

In order to find out this boundary, scientists tried to do double- slit experiment with larger molecules to see how large the particle size would cease the interference pattern. Arndt's group sent the molecule of more than 800 atoms into a double- slit and they got the interference pattern [7]. The search for the threshold continues. Recently, the research team sent hot complex molecule consisting of nearly two thousand atoms with a mass greater than 25,000 atomic mass units into the double slit, and certainly obtained interference pattern [8]. It is worth mentioning that the complex massive molecule beam sent to the double slit is very hot, but the de Broglie wavelength for such massive molecule is only about 53 fm, which is five orders of magnitude smaller than the diameter of the molecular itself. An essential question must then be asked: what mechanism prevents larger objects from existing in superposition state, forcing the separation of the quantum world from the classical world.

We already know that quantum world results from the random motion of subatomic particles. The degree of random motion of subatomic particles depends upon the root mean square speed $\upsilon_{rms} = (\frac{3kT}{m})^{1/2}$, which is proportional to the square root of the ambient temperature and reversely proportional to the square root of the particle's mass. Table 1 shows that at a given temperature, the random-motion speed of a particle decreases with increasing its mass, which leads to heavier particle approaching the classical world.

The $\upsilon_{rms}$ speed of silicon particle with $1.4 \times 10^5$ atomic mass units is as low as 7.2 m/s at room temperature and 15 m/s at 1000°C, calculated by equation 2. Obviously, under room temperature, it will become extremely hard to get an interference pattern by sending that massive particles to a double slit system. Since the speed ($\upsilon_{rms}$) of the particle is too low to overcome the effect of gravitational force on its random motion. The calculation result of Equation (2) shows that the speed $\upsilon_{rms}$ of the complex molecule with mass greater than 25,000 atomic mass units used by Markus Arndt's team is about 17 m/s at room temperature and 35 m/s at temperature up to1000$^o$C. Clearly, this complex molecule is already close to the room temperature threshold, while there is still a little room for the high temperature double slit interference experiment. In contrast to the high temperature double slit experiment, if the temperature of environment in which the subatomic particles move becomes absolute zero, and without any other disturbing field, the subatomic particles will cease making any random motion in free space, and all subatomic particles must obey the physical laws of classical mechanics.

Now we reach the conclusion that there is no a fixed boundary between quantum and classical worlds. The boundary shifts with the particle's random motion speed. The higher the temperature, the more massive particle will become the member of quantum world. While at absolute zero temperature, almost all free particles will move in classical world.

However, the subatomic particles in confined system, like electrons in atoms never cease playing their random motion even at absolute zero temperature. Therefore, particles in a confined system always belong to quantum world. Now, we can see that in addition to the random motion of subatomic particles, squeezing particles into much smaller confinement system is another necessary condition for achieving quantum mechanics.

In contrast with the effect of confinement on the particle's physical properties, the speed of particles under an external field also can change their intrinsic properties. Let us suppose that a particle moves with the speed hundred times higher than its $\upsilon_{rms}$ speed, then physical properties of the particle must be described by classical mechanics, and the energy associated with the random motion of the particle can be considered as a perturbation. Under the modulation of random motion, the particle moving trajectory should show an irregular wave pattern. In contrast to the case of de Broglie waves, the average wavelength of a moving particle increases with the particle's translating speed. Similarly, under the action of an external field, electrons in conductors and semiconductors exhibit moving trajectories in random walk pattern, or move with their wave function. It is has been proved that the electrons in superconductors move with trajectories in a regular wave pattern [9].

Hence, it is a natural conclusion that excepting in superconductors, electron always manipulates themselves as particles moving with trajectories in random walk patterns, rather than that electrons have particle - wave dual nature, or that one electron can occupy two places at the same time. On the other hand, if particles move with a speed much lower compared with its random motion speed $\upsilon_{rms}$, the particle acting like a wave packet or wave function moves and diffuses with evolvement of time, which is the exact form happened in double slit interference experiments.

## IV  Double Slit Interference Experiment

The double slit experiment is a touchstone for testing the correctness of quantum mechanics. It is usually adopted to discuss the superposition principle of quantum mechanics, and the wave nature of material particles. A large number of double- slit experiments have been performed with matter particles, including electrons, atoms, molecules, $C_{60}$ fullerenes and massive particles, even up to date, the massive complex molecules composed of 2,000 atoms as mentioned above [8]. So far, the conclusion already reached for all of quantum particles showing the interference pattern in double slit experiment is that the wave nature of the matter particles makes a particle to pass through both slits at the same time, and interference with themselves on the screen. This baffling conclusion is too counterintuitive to accept. We have already emphasized above that as long as the scientific conclusions sharply conflict with the laws of nature, some important causality must be ignored, because the cause and effect laws govern the entire development of science.

Let us suppose that the electron gun fires one electron at a time towards the double slits, and the firing speed is not too high compare to the random motion speed of the electrons. In this case, the electron behaving as a wave packet or a wave function moves with the firing speed and at the same time diffuses outward with time evolving. The schematic diagram in Figure 2a shows a two dimensional electronic double - slit experimental apparatus.

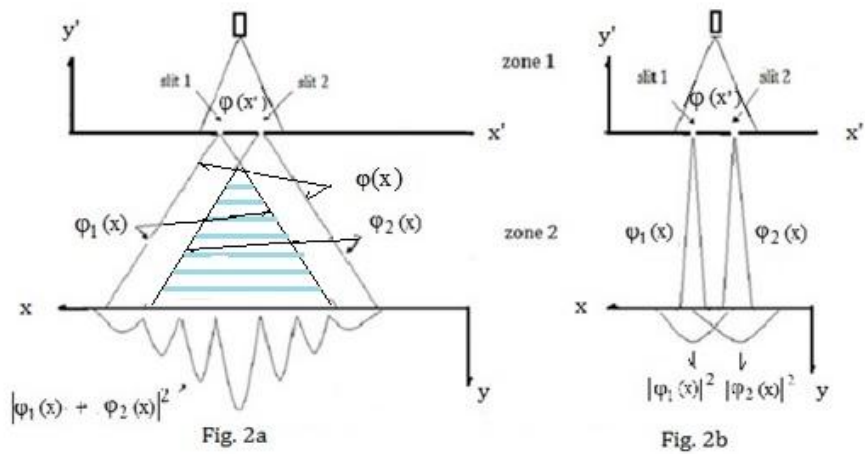

Fig. 2a

Fig. 2b

Fig 2. A schematic diagram of a double-slit experiment with electrons. The electronic gun at top shoots one electron at a time to the double-slit apparatus. The initial wave function of the shot electron in zone 1 is $\varphi(x')$ in $(x', y')$ coordinate system, which passes through slit 1 and slit 2 forming wave function $\varphi_1(x)$ and $\varphi_2(x)$ in zone 2, respectively. The initial wave function $\varphi(x')$ becomes $\varphi(x)$ in $(x, y)$ coordinate system. The electron passing through any one of the double slit will fall into the superposition state $\varphi(x) = \varphi_1(x) + \varphi_2(x)$, which will create a interference pattern on the screen as shown in Fig. 2a. If the wave function $\varphi_1(x)$ and $\varphi_2(x)$ do not overlap each other before reaching to the screen, then interference pattern will be lost, there are only two peaks, one comes through slit 1, $|\varphi_1(x)|^2$, the other one comes through slit 2, $|\varphi_2(x)|^2$ as shown in Fig. 2b.

In Figure 2a, we use $\varphi(x')$ to represent the wave function of the electron fired to the double slit in $(x', y')$ coordinate system. The centers of slits 1 and 2 are located at $x_1'$ and $x_2'$, respectively. There is only one electron in the initial wave function $\varphi(x')$ at each time. In the two dimensional diagram of double- slit experimental apparatus, before the electron passes through the two slits, the initial wave function $\varphi(x')$ is located in the $(x', y')$ coordinate, zone 1, as shown in Figure 2a. The probability of the electron passing through slit 1 and slit 2 should be $|\varphi(x_1')|^2 \cdot \Delta$ and $|\varphi(x_2')|^2 \cdot \Delta$, respectively, $\Delta$ is the width of both slit. The wave functions of the electron passing slit 1 and slit 2 are expressed as $\varphi_1(x)$ and $\varphi_2(x)$ in zone 2, respectively. After the electron passed through either one of the two slits, the initial wave function $\varphi(x')$ becomes $\varphi(x)$ in (x, y) coordinate system, as shown in Figure 2a. Assuming that the interaction between electrons and slits is the elastic, then both wave functions $\varphi_1(x)$ and $\varphi_2(x)$ have the same energy state as the initial wave function $\varphi(x)$, therefore the two wave functions, $\varphi_1(x)$ and $\varphi_2(x)$, are in the superposition state of the initial wave function $\varphi(x)$, that is

$$\varphi(x) = \varphi_1(x) + \varphi_2(x) \tag{11}$$

If the electron has a relatively high random motion speed, then the two wave function $\varphi_1(x)$ and $\varphi_2(x)$ have a large overlap between them (the shaded part as shown in Figure 2a). Thus, the electron passing through slit 1 can also partially stay in wave function $\varphi_2(x)$. Similarly, the electron passing through slit 2 can also partially stay in wave function $\varphi_1(x)$. This means that no matter which slit the electron traverses through, they must stay in the same superposition state $\varphi_1(x) + \varphi_2(x)$. Since the probability of detecting an electron on the screen is governed by the square of superposition wave function, and the probability detecting an electron at position x can be expressed as

$$|\varphi(x)|^2 = |(\varphi(x)_1 + \varphi(x)_2)|^2 \text{ or}$$

$$|\varphi_1(x)|^2 + |\varphi_2(x)|^2 + 2|\varphi_1(x)||\varphi_2(x)|$$

Now we can see that the interference pattern obtained by sending one electron at a time should be the same as the interferogram obtained by sending a large number of electrons through the double slit. They are all controlled by the same superposition state. Here $|\varphi_1(x)|^2$ is the probability of detecting an electron at the screen when slit 1 is open and slit 2 is closed. Similarly, $|\varphi_2(x)|^2$ is the probability for detecting an electron at the screen as slit 2 is open and slit 1 is closed. The term $|\varphi_1(x)||\varphi_2(x)|$ is so called the interference term, which dominates the interference pattern.

Generally speaking, if a particle's random-motion speed at room temperature is not high enough, then the two wave functions of the particles passing through two slits cannot overlap each other, the interference term will become zero, resulting in two symmetric peaks around each slit's center, shown in Figure 2b. For instance, Arndt's complex molecule composed of nearly two thousand atoms has a random motion speed about 17 m/s at room temperature, which seems difficult to achieve an interference pattern in a double-slit experiment. If the molecule is heated to thousand degrees Celsius, then the molecule's random motion speed will become 35 m/s calculated by the equation 2, in this case it seems possible to achieve an interference pattern.

Excepting for the effect of temperature on interference pattern, the firing speed of the electron also plays an important role in double slit experiment. Juffmann found that when the molecules travel through double slit with a high velocity, the interference pattern disappears [10]. Olaf Nairz et al performed a beautiful quantum interference experiment with $C_{60}$ fullerenes [11]. In their double-slit experiments, all $C_{60}$ fullerene molecules were heated to an internal temperature in excess of 3,000° K by the laser beam. In this case, the $\upsilon_{rms}$ speed calculated from equation (2) can reach about 320 m/s. When the hot $C_{60}$ fullerene is sent to double slits with a mean velocity 200 m/s, the interference pattern only showed the first order peak; while the hot $C_{60}$ fullerene is sent with a lower mean velocity 117 m/s, the interference pattern is developed to show third order interference peak. They also observed that the sent wave packet has experienced spread out during the evolution process. The reason is that when the particle is sent to the double slit at a very high velocity, the time interval of the particle passing through the distance between slits and screen is too short to form sufficient overlap between the wave functions passing through the both slits.

The another big puzzle of the double-slit experiment is that the electrons cease creating an interference pattern when a detector is set up near one of the slits to determine which slit an electron is passing through. Under this situation, the electrons acting as classical particles will simply create two peaks as shown in Figure 2b. We have mentioned above that if the interaction between the incident electrons and slits is the elastic one, the two wave functions passing through the two slits have the same energy state as the initial incident wave function. Therefore, the interference of the two wave function $\varphi_1(x)$ and $\varphi_2(x)$ satisfies the superposition principle of the initial wave function $\varphi(x)$. However, if detectors are placed near one slit or both slits, to see which slit an electron is passing through. That will induce an inelastic interaction between the electrons and detector, and lead to a change in the energy of the electron being determined. Therefore, the wave function of the electron interacting with the detector does not have the same energy state as the initial wave function. Thus the two wave functions $\varphi_1(x)$ and $\varphi_2(x)$ do not satisfy the requirement of quantum superposition principle. So that, there is no available superposition state between these two wave functions. Although, the two function, $\varphi_1(x)$ and $\varphi_2(x)$ may still have some overlap each other, but the interference term must be zero. The excellently experimental results published by S. Frabboni et al. have proved that the detector placed near slits can change the type of scattering; the elastically scattered electrons can construct an interference pattern, while the inelastic scattered electrons have no ability to create the interference pattern [12]. All of facts above further prove that double-slit experiment is a touchstone for testing the correctness of quantum mechanics, which was selected as the "most beautiful experiment in physics"; it is the core of quantum mechanics.

On the other hand, scientists have tried double-slit experiment using large size and massive (heavy) particles, and noted that these large and massive particles have the correspondingly very small de Broglie wavelength. For instance, Stefan Girlish et al [13] performed quantum interference experiment on organic molecules consisting of 430 atoms, the mass can reach 6910 amu, the largest size of the organic molecules is up to 60 Å, and the de Broglie wavelengths down to 1 pm. In addition, in Naiiz's interference experiment [11], the de Broglie wavelength of $C_{60}$ fullerene is as low as 2.8 pm, which are five orders of magnitude smaller than the particle size (0.71 nm, diameter). However, in the Markus Arndt's interference experiment, a complex molecules of 2,000 atoms with a mass of 2.5 x $10^4$ amu has the de Broglie wavelength as small as 53 fm, which is five orders of magnitude smaller than the diameter of the molecular itself.

Based on the model of quantization of angular momentum in the Bohr atoms, the de Broglie wave should be the particle wave, that is, the moving trajectory of a particle shows a wave pattern. The energy and mass of a particle always remain in the same entity within its moving trajectory. The above experiment results submit a question; How can such extremely small de Broglie wavelength (five orders of magnitude smaller than the diameter of the molecular itself) make the relatively larger particle occupy the two slits at the same time in a double slit experiment? The answer is absolutely impossible.

Now we can conclude that the interference pattern created by quantum particles in the double-slit experiment must also results from the random motion of quantum particles, not from the so-called particle- wave nature originated from the de Broglie wave [14].

## V. On Quantum Entanglement

Another of the strangest aspect of quantum mechanics is so- called quantum entanglement. Entanglement occurs when the quantum states of two or more particles are linked together; even these entangled particles are not physically connected. Thus, even if the particles are separated by a large distance, the quantum state of each particle cannot be described independently of the states of the other particles. As a result, the measurement outcomes of one particle must be highly correlated with the measurement outcomes of other particle. The idea of quantum entanglement originated in thought experiment of Einstein, which aimed to emphasize that Heisenberg's uncertainty principle is incorrect and the quantum mechanical description of physical reality by wave function is incomplete. In the well-known EPR article, the authors supposed that we could produce two electrons whose momentum and positions are linked [15]. After the two electrons move apart, they are matching correlation between both the positions of the two electrons and their momentum. We can measure the momentum of the first one and know the momentum of the second one, but we could also measure the position of the second one, and then know the position of the first one.

Obviously, the two linked electrons described in EPR article are classical electrons. This thought experiment clearly reflects the fact that the founders of quantum mechanics truly believed that quantum particles, like electrons, must strictly obey the laws of classical physics and cannot have any random movement. The probabilistic nature of quantum particles is caused by the process measurement process. Some of the double-slit interference experiments mentioned above demonstrate that the core nature of quantum mechanics is the random motion of quantum particles. At normal temperature, the two linked electrons move apart and spread out over time like two wave functions. The physical quantities such as positions, momentum, and spins for each electron in the two wave functions do not have the fixed correlation, owing to the random motion of electrons in the two wave functions. Therefore, measurements of position or momentum on a particular electron will not allow prediction of measurement outcomes of position or momentum on other electron. However, the energy states of the two electrons do have a physical connection. They all move in each other's long region potential field ($\frac{e}{r_{12}}$), as the distance $r_{12}$ between the two entangled electron increases, the potential field becomes weaker and the entanglement of the electron pair decays spontaneously. More precisely, for quantum particles that are always in random motion in a linked system, it is impossible to alter instantaneously the properties of distant particles by acting on local particles. The conclusion is that the quantum particles that satisfy the Schrodinger equation should not exhibit any entanglement effect. As we mentioned earlier, photons do not belong to the family of quantum particle that satisfy the Schrödinger equation, so the properties of photons cannot be described by quantum mechanics. In other words, the entanglement effect of photons cannot be applied to quantum particles that satisfy the Schrodinger equation.

## VI  Matrix Mechanics and Wave Mechanics

In the early 20[th] century, Planck postulated that matter could only radiate light in discrete portions of energy, which he called quanta, through studying blackbody radiation curves. That is, as if light is composed of particles, then light of frequency f can emit energy E=hf. In 1905, Einstein applied Planck's postulation to explain successfully the photoelectric effect, and postulated that the maximum kinetic energy of the liberated electron must be $KE_{max} = hf - \Phi$, where $\Phi$ is the work function.

In 1913, in order to obtain a discrete set of stable orbits, Bohr postulated without any proof that the quantized values of circular orbital angular momentum, L= nh/2π, where h is Planck constant and n is an integer. Later, Sommerfeld extended Bohr's quantization rule to include elliptical orbits, this led to the so-called Bohr-Sommerfeld quantization rule

$\oint p \delta q$ = nh/2π

Where, q is the canonical coordinate, and p is the momentum, $\oint$ denotes the line integral around a closed orbit (old quantum condition). No evidences above can be explained in the laws of the classical physics and all equations contain the quantum parameter h.

Faced with this situation, physicists at that time believed in common that quantum mechanics could be realized if the Hamiltonian equation of classical mechanics was modified suitably to introduce h into the equation.

Matrix mechanics, in fact, originates from Heisenberg's quantum condition. Heisenberg assumed that the electron still obeyed the classical equations of Newton, but replaced the electron's position coordinate with some quantum quantities, since be believed that the position of the electron was non-observable. In the case of one degree of freedom, the Bohr-Sommerfeld's quantum condition is:

$\oint m\dot{x} \, dx$ = nh         (12)

In a one-dimensional periodic system, the motion of an electron can be described by the time-dependent position $x$ (n, t), which can be represented by a Fourier series

$x$ (n, t) = $\sum_\alpha a_\alpha(n) e^{i\alpha\omega(n)t}$         (13)

Where n is the number of stationary states, first integrating equation (12) over the entire period of motion, then substituting for $x$ with the Fourier series (13), and differentiating with respect to n, gives:

$$h = 2\pi m \sum_{\alpha=\infty}^{\infty} \alpha \frac{d}{dn} |a_\alpha(n)|^2 \alpha \omega(n) \qquad (14)$$

Then Heisenberg converted equation (14) to

$$h = 4\pi m \sum_{\alpha=0}^{\infty} \{|a(n+\alpha, n)|^2 \omega(n+\alpha, n) - |a(n, n-\alpha)|^2 \omega(n, n-\alpha)\} \qquad (15)$$

This is Heisenberg's famous quantum condition. Heisenberg assumed that $|a(n, n-\alpha)|^2$ is proportional the transition probabilities from n to n−$\alpha$ state [16].

After Heisenberg completed his paper, he handed it over to Born to decide whether it was worth publishing (Born was Heisenberg's supervisor at the time). After reading Heisenberg's paper, Born wrote" I am deeply impressed by Heisenberg's manuscripts, which is a great step forward in the program we have been pursuing …. . I suddenly see light; Heisenberg's symbolic multiplication is nothing but a matrix calculation, well known to me. Instead of $a$ (n, n+ $a$), I wrote q (n, m), and rewrote Heisenberg's form of Bohr's quantum conditions… I mean that the two matrix products of **pq** and **qp** are not identical. I am familiar with the fact that matrix multiplication is not commutative. …..closer inspection shows that Heisenberg's formula gives the values of diagonal elements (m = n) of the matrix **pq − qp**; it says that they are all equal and have the value h/2πi. But what are the other element m ≠ n? Repeating Heisenberg's calculation in matrix notation, I soon convinced myself that the only reasonable value of the non-diagonal elements should be zero, and I wrote down the strange equation

$$\mathbf{pq - qp} = \frac{h}{2\pi i} \mathbf{1} \qquad (16)$$

Where **1** is the unit matrix. But this is only a guess, and my attempts to prove it have failed" [17]. Later, Born's student Pascual Jordan proved that all nondiagonal elements of **pq − qp** are equal to zero. Then in 1925, the non-commuting equation was emerged [18]. Physicists at the time regard it as a fundamental law of quantum mechanics.

Since the non-commuting equation (16) is a diagonal matrix, which satisfies the calculative rules of Poisson bracket of classical mechanics, as it follows the equation of motion of classical mechanics in an one-dimensional quantum system

$$\dot{\mathbf{q}} = \frac{\partial H}{\partial p} \text{ and } \dot{\mathbf{p}} = \frac{\partial H}{\partial q} \qquad (17)$$

The equation of motion in matrix mechanics can be rewritten as

$$\dot{\mathbf{q}} = \frac{2\pi i}{h} (\mathbf{Hq - qH}), \quad \dot{\mathbf{p}} = \frac{2\pi i}{h} (\mathbf{Hp - pH}) \qquad (18)$$

In order to make the abstract equations more concrete, the equations of motion of a free electron in one- dimension ($\hat{H} = \frac{\hat{p}^2}{2m}$) are given below, derived by Achim Kempf [19].

$$\hat{x}(t) = \begin{pmatrix} 0 & \sqrt{1}\left(L - \frac{ih(t-t_0)}{2Lm}\right) & 0 \\ \sqrt{1}\left(L + \frac{ih(t-t_0)}{2Lm}\right) & 0 & \sqrt{2}\left(L - \frac{ih(t-t_0)}{2Lm}\right) \\ 0 & \sqrt{2}\left(L + \frac{ih(t-t_0)}{2Lm}\right) & 0 \\ & & & \ddots \end{pmatrix} \quad (19)$$

And

$$\hat{p}(t) = \begin{pmatrix} 0 & -\sqrt{1}\,\frac{i\hbar}{2L} & 0 \\ \sqrt{1}\,\frac{i\hbar}{2L} & 0 & -\sqrt{2}\,\frac{i\hbar}{2L} \\ 0 & \sqrt{2}\,\frac{i\hbar}{2L} & 0 \\ & & & \ddots \end{pmatrix} \quad (20)$$

Where $t_o$ is the initial time, m is the electron mass, L is some arbitrary real number with units of length. The matrix equations of motion are shown in an indefinite matrix representation.

Through complete analogy with classical Hamilton-Jacobi equation, the founders found that the energy levels in a quantum system could be derived by transforming the Hamiltonian matrix H to its diagonal W using a transformation matrix S,

$$H(q, p) = S\,H(q_o, p_o)\,S^{-1} = W \quad (21)$$

This is the energy equation of matrix mechanics, published by Born, Heisenberg and Jordon, so-called the three man's paper in 1926 [20]. As can be seen from the one-dimensional motion equation for free electrons given above, this energy equation is extremely difficult to be handled. In the same year, W. Pauli used this energy equation to derive the correct energy spectrum of the hydrogen atom [21]. From then on, no one, including the authors, has used this equation to study any quantum system. Thus, the correctness of this equation has not been completely proved.

Based on the non-commutative relation between quantum particle's position and its momentum, Heisenberg proposed uncertainty principle in 1927:

$$\Delta p \Delta q > h \quad (22)$$

This means that the more accurately the position is determined, the less accurately the momentum is known, and vice verse **[22].**

Now, we can summarize that the non-commuting equation (16), the equations of motion (18), the energy equation (21), and Heisenberg's uncertainty principle (22), all together constitute the backbone of matrix mechanics.

Obviously, matrix mechanics is built on the classical Hamiltonian equation, which is modified based on Bohr- Sommerfield old quantum condition (orbital angular momentum, L= nh/2π). Except the uncertainty equation, all other equations like classical physics are non-linear deterministic equation. That is, the moving trajectory of the quantum particle is completely governed by the equation of motion. There is a fixed relation between a particle's position and its momentum. However, due to the uncertainty relation, their real values at a given point of the moving trajectory cannot be accurately measured, thereby leading to the probabilistic results of the deterministic equations in matrix mechanics. This is as Heisenberg explained in his 1927 work: Canonically conjugate quantities can only be determined simultaneously with a characteristic inaccuracy. This inaccuracy is the actual reason for the occurrence of statistical relationships in quantum mechanics [22].

In matrix mechanics, both the equations of motion (18) and energy equation (21) are deterministic and continuous equations. In terms of Heisenberg's quantum condition (22), some certain states are selected as allowed states from the continuous energy. In fact, the Heisenberg's quantum condition is just the Bohr - Sommerfeld quantization condition in the matrix representation. The Bohr-Sommerfeld quantum condition seems to be an acceptable assumption in the development history of quantum mechanics. Strictly speaking, however, the Bohr-Sommerfeld quantum condition does not hold in modern quantum theory, because the Schrodinger equation does not produce well-defined orbits for electrons. The wave function only gives the electronic cloud that describes the distribution of the electron in different state. Thus, in principle, matrix mechanics should belong in the realm of the old quantum theory.

In 1926, Schrödinger published his equation in series papers [23]. The one dimension time-dependent Schrödinger equation is:

$$i\hbar \frac{d\varphi(x.t)}{dt} = -\frac{\hbar^2}{2m}\frac{d^2}{dx^2}\varphi(x.t) + V\varphi(x.t) \qquad (23)$$

This was called the wave mechanics. We have demonstrated in section II of this article that the essential nature of quantum particles is their random motion, which means that there is no any fixed relation between particle's position and its momentum. We proposed that a reasonable path to derive this equation might lie in applying the classical wave equation to describe the physical system in which particles move randomly. Therefore, the reality described by the wave function is the quantum particle in random motion.

The Schrödinger equation does not contain particle's coordinate variable q and momentum variable p, which hide in its wave function. That is, they belong to immeasurable physical quantities in wave mechanics. It follows from this that there is no need of the non-commuting relation, and the uncertainty relation in wave mechanics. Because we only know the probability that a quantum particle located at a given place, how can the particle's position and its momentum be measured precisely? Thus, it is meaningless to measure precisely these immeasurable physical quantities.

So we are forced to conclude that the physical realms of matrix mechanics and wave mechanics are in conflict with each other. We only can accept one of them, not a mix, or we will never understand quantum mechanics.

# VII  Heisenberg's Uncertainty Relations

Since the uncertainty relations were published, the debates over the validity, reasoning and foundation of the uncertainty relations have never ceased throughout the entire history of quantum mechanics. Heisenberg wrote in his paper, "not only the determinism of classical physical must be abandoned, but also the native concept of reality, which looked upon the particles of atomic physics as if they were very small grains of sand…… if its position is determined with increasing accuracy, the possibility of ascertaining the velocity becomes less, and vice versa" [22, 1]. To carry this argument one-step further, Max Born, one of the founders of quantum mechanics, more precisely wrote in his famous paper "the Statistical Interpretation of Quantum Mechanics", "that in short, ordinary (classical) mechanics must also be statistically formulated….it is possible without difficulty… I should like only to say this: the determinism of classical physics turns out to be an illusion, created by overrating mathematics ---logical concepts. It is an idol, not an ideal in scientific research and cannot therefore be used as an objection to the essentially in deterministic statistical interpretation of quantum mechanics'' [1].

From the viewpoints of the above two founders, it is obvious that the quantum uncertainty relation should be the very foundation law that both the quantum world and the classical world should obey. However, Einstein never accepted the uncertainty relations and tried his best to get physics rid of the uncertainty relations. Einstein believed that the position and momentum of a particle existed objectively apart from how we measure them. Einstein wrote, **"**Like the moon has a definite position, whether or not we look at the moon, the same must also hold for the atomic objects. . . . Observation cannot create an element of reality like a position, there must be something contained in the complete description of physical reality which corresponds to the possibility of observing a position, already before the observation has been actually made" [24]**.** In order to defend the foundation of science, Einstein tried with all his might to think out methods of measurement by means of which the position and momentum of a quantum particles can be measured as accurately as the measuring apparatus allowed. Undergoing the lifetime debate between Einstein and the Copenhagen team, unfortunately he lost the game but held the truth.

The essential nature of the uncertainty relations is to deny the conservation laws of both energy and momentum. However, the conservation laws lay the foundation for all science. Without these two foundation laws, nature would become unpredictable and everything in nature must be controlled by uncertainty.  After the uncertainty relation became the foundation of quantum mechanics, a number of counterintuitive concepts in quantum mechanics undoubtedly arise from the uncertainty relations. For instance, quantum particles like ghost have no position and no speed, but can occupy two places at the same time. In addition, uncertainty relations allow violation of conservation laws of energy and momentum, so that**;** electrons can borrow energy to climb over a barrier for very short period, and energy of a quantum system can be so highly uncertain that the particles can emerge from empty space, and so on. All these unscientific beliefs have become the foundation of quantum mechanics.

In the final analysis, all of debates on quantum mechanics concentrate at last on the central question: what is the essential nature of quantum particles? If quantum particles are treated as classical particles having fixed relation between a particle's position and its momentum as matrix mechanics does, any debate will not lead to a unified conclusion. Instead, if we accept the fact that quantum particles always move randomly, there is no any fixed relation between particle's position and its momentum, and then all debates have no reason to stay.

In wave mechanics, however, for a given energy state, the particles move randomly in their wave function, there is no any fixed relation between the particle's position and its momentum, and they can take any value allowed by boundary conditions without affecting the energy state. The position and momentum of quantum particles can be regarded as the hidden variables in their wave function, which should belong to immeasurable physical quantities. Even if at a given time, the exact values of the particle's position and its momentum were known with certainly, this would have nothing to do with determining energy state of the particle. Thus, it is meaningless to measure precisely a particle's position and its momentum at a given moment. Therefore, there is no need for the non-commuting relation and Heisenberg uncertainty relations in quantum mechanics.

Because the Heisenberg's uncertainty relation was built on the foundation of matrix mechanics, so it was designed for classical particles rather than quantum particles. For a classical particle without in any way disturbing it, no matter how small it is it must obey the classical Hamiltonian equation; at any moment of time, its position and momentum can be precisely predicted and measured. The classical world has known for a long time that measuring objects disturbs them, so called observer effect, but in principle, the disturbance can be reduced infinitely. We believe that the fundamental limitation on the accuracy of measuring physical quantities lies in the technological realm, not in the physical principle realm. With development of science and technology, scientists can avoid the disturbance. Therefore, the scientists can measure things with the degree of accuracy as high as they want, and take themselves out of their measurements.

Heisenberg did not provide a general derivation of his uncertainty relations. In fact, he even did not give a precise definition for the uncertainties Δq and Δp; instead, his uncertainty relations were based on some estimates. In Chicago lecture, he gave his refined uncertainty relations (22) [25].

In 1929, two years later after Heisenberg published his uncertainty relations, Roberson gave a more general mathematical derivation based on the non-commutative relation of p and x leading to other uncertainty relations as

Δx Δp ≥ $\hbar$/2 and ΔE Δt ≥ $\hbar$/2

There $\hbar$ is the reduced Plank constant, $\hbar = h/2\pi$. The Roberson uncertainty relations are generally accepted as the standard one instead of Heisenberg's original formulas (22) [26].

In 2003 Masanao Ozawa at university of Nagoya, Japan, recalculated the measurement-disturbance relation and generalized the Heisenberg's uncertainty relation to a new formula that holds for all possible quantum measurement. The Ozawa's formulation would also disturb the system far less than Heisenberg's formulas [27].

In 2012, the team at university of Toronto set up an apparatus to measure the polarization of entangled photons [28]. The team's main goal was to quantity how much the operation of measuring the polarization disturbs the photons. By comparing thousands of measurement results, the authors revealed that their precise measurements disturbed the system much less than the original Heisenberg's formula predicted. The results provided the evidence that Ozawa's formula is more accurate.

In 2014, S. Sponar et.al reported a neutron-optical experiment that records the error of a spin-component measurement, as well as the disturbance caused on a measurement of another spin measurement to test error-disturbance uncertainty relation [29]. Their experimental results demonstrated that Heisenberg's original uncertainty relation is not correct, and the Ozawa and Sponar formula are valid in a wide range of experimental parameters.

It seems to us that the criterion for judging the correctness of the Heisenberg's uncertainty relations is not lying in how to make refinements of the mathematical methods, but in how to understand correctly the fundamental concepts of quantum mechanics.

In accordance with the discussions on quantum mechanics above, we can draw a conclusion that as long as the particle does not do random motion and there is a fixed relation between its two physical quantities described by the non-commutation operator, then the particle belongs to the classical world.

For further consideration, we need to re-examine Heisenberg's microscope thought experiment. Based on his thought experiment, a localized electron was observed by using γ-ray microscope; Heisenberg derived his uncertainty relation of position and momentum by means of the theory of the Compton Effect. Then, he could argue that the measurement precision of the position of an electron and its momentum could dominated by his uncertainty relation [22]. Obviously, the localized electron observed by Heisenberg's microscope is a typically classical particle with no any random motion. The interaction between a photon and the localized electron can be described by the well-known Compton scattering formula as below

$$\mathcal{E}' = \frac{\mathcal{E}}{1 + \mathcal{E}/m_e c^2 (1 - \cos\theta)}$$

There $\mathcal{E}$ and $\mathcal{E}'$ are the photon energies before and after interaction, respectively. $\theta$ is the angle at which the photon is deflected in encounter. Here, $m_e$ is mass of electron, c is the speed of light. We can see from the equation that if the deflected angle θ is set to zero in Heisenberg's thought experiment for viewing electron's position, then the energy of the incident photon equals the energy of the scattered photon. That is, the interaction between the incident photon and the measured electron is elastic. The incident photon does not transform any of its energy and momentum to the measured electron. Thus, the disturbance caused by measuring electron's position can be unlimited small. Theoretically, the lower bound of the product of error and disturbance in measuring a classical particle can be arbitrarily small.

Mathematically, Roberson derived the uncertainty relation from the non-commuting relation based on the standard deviation equation. For example, a pair non-commuting operator $\hat{A}$ and $\hat{B}$, the product of the standard deviation ΔA and ΔB obey the following inequality

$$\Delta A\, \Delta B \geq \frac{i}{2}\, |<\varphi[\hat{A},\hat{B}]\varphi>|$$

There, $[\hat{A},\hat{B}]$ stands for the commutator, and $[\hat{A},\hat{B}] = \hat{A}\hat{B} - \hat{B}\hat{A}$, φ is the wave function. The standard deviation is defined as

$$\Delta A = (<\varphi, \hat{A}^2\varphi> - <\varphi, \hat{A}\varphi>^2)^{1/2}$$
$$\Delta B = (<\varphi, \hat{B}^2\varphi> - <\varphi, \hat{B}\varphi>^2)^{1/2}$$

In addition, the wave function φ is assumed normalized. If the commutator is a constant, as in the case of conjugate operators $[\hat{A}, \hat{B}] = -i\hbar$, then we obtain the following equation,

$$\Delta A\, \Delta B \geq \frac{i}{2} [\hat{A},\hat{B}]$$

Replace ΔA, ΔB with Δx, Δp, the standard deviation of position and momentum in one dimension, we get the Heisenberg's relation

$$\Delta x\, \Delta p \geq \hbar/2$$

Mathematically, the root cause of the uncertainty relations lies in the non-commuting relation of operators [26].

In quantum mechanics, it is generally assumed that any observable physical quantity that can be measured in a physical experiment should be associated with a self-adjoint operator. The operator must yield real eigenvalues, since they are values that can be the result of corresponding experiment. Based on the proposed reality of Schrödinger equation in this paper, Schrödinger equation describes a system containing random-motion particles, the particle's position and its momentum should belong to immeasurable physical quantities. There should be no existing operators associated with these immeasurable physical quantities. Because the coordinate variable x in the Schrödinger equation of one-dimension is used to describe the wave function, rather than the coordinate variable of the internal random-motion particle. In quantum mechanics, it is long-term misunderstanding to regard the coordinate variable x of the wave function as the variable of a random-motion particle.

Furthermore, in quantum mechanics, the position operator of a particle in one dimension is represented as $\hat{x}$. When the position operator acts upon wave function, we then have the formula $\hat{x}\varphi(x) = x\,\varphi(x)$, where the independent variable x does not in any way represent the particle's position. According to the definition of operators, the average value of the so called "particle's position" should be $<\varphi(x)|x|\varphi(x)>$, leading to a weird result that was found to have a dimension of area (x squared). Therefore, the position operator in quantum mechanics does not have an objective reality associated with it. In addition, it is no any physical meaning to find out the average value of an independent variable.

The key important point is that because the variable x is used to describe the wave function, it is independent of the wave function. However, if the independent variable x is defined as a position operator, then the independent variable x must be related to the wave function. Hence, we can see that it is contradictory to treat the variable x as an independent variable and as a positional operator. Thus, it is reasonable to believe that in the history of quantum mechanics, a major conceptual mistake made by the founders was to regard the coordinate and time of the independent variables of the wave function as operators. It is just this fundamental mistake, which opened the door for uncertainty relations and other irrationality things to enter quantum physics.

Based on the discussions above we can say that in wave mechanics, both the particle's position and its momentum belong to immeasurable physical quantities. Thus, there are no existing operators and non-commutation relation associated with them. If the non-commutation relation $[\hat{p}, \hat{x}] = -i\hbar$ does not hold, then the Heisenberg's uncertainty relation of position and momentum has no mathematical basis to support it staying in the quantum mechanics.

Heisenberg also defined a time operator and used the non-commutation relation of time-energy relation $[\hat{E}, \hat{t}] = -i\hbar$ obtaining a time- energy uncertainty relation, $\Delta E \Delta t \geq h$. Here $\Delta E$ is the accuracy of energy measurement and $\Delta t$ can be attributed to either the accuracy or the duration of a measurement. If say the uncertainty relation of position-momentum has some physical reason to make people to accept it, then the uncertainty relation of time-energy is just a claim based on his belief without any reliable scientific justification. Pauli once gave a powerful argument against the existence of a time operator, he wrote "we conclude that the introduction of a time operator must be abandoned fundamentally, and that time t in quantum mechanics has to be regarded as an ordinary real number" [30]. Pauli, like Einstein, also did not accept the uncertainty relation of position-momentum. Pauli wrote, "The p´s must be assumed to be controlled, the q′s uncontrolled, the more you learned about one, the less you could say about the other. The physics of this is unclear to me from top to bottom. The first question is why only the p´s (and not simultaneously both the p´s and the q´s) can be described with any degree of precision". Pauli summed it up as follows, you can look at the world with the p-eye or with the q-eye, but open both eyes together and you go wrong. What could this mean?" [31]

It is generally accepted that the definition of time is the same in both classical and quantum world. Time is measured by substance motion, and it becomes evident through the motion. If time were regarded as an operator, then what would happen as the time operator acting on the wave function? By the definition of operators, the average value of time becomes time squared. Obviously, the theory is not self consistent. Based on the evidences given above, time cannot be defined as an operator of quantum mechanics. From this follows that the non-commutation relations, such as $[\hat{p}, \hat{x}] = -i\hbar$ and $[\hat{E}, \hat{t}] = -i\hbar$, are all man-made, there are no such kind things in wave mechanics.

Einstein argued in the EPR article [15] that "If the operators corresponding to two physical quantities, say A and B (p and q) do not commute, then the precise knowledge of one of them precludes such knowledge of the other. From this follows that the quantum mechanical description of reality gives by the wave function is not complete." Einstein's conclusion is based on assuming that the non-commutation relation is correct, then leading to the conclusion that the wave function is not complete. However, innumerable experimental results demonstrate that the wave function is the most complete description of a physical system. Based on Einstein's arguments, if we accept that the wave function of Schrodinger equation can certainly give the most complete description to a physical system, and then the non-commutation relations in quantum mechanics must be abandoned.

## VIII Discussion

In this section, we will prove that in both the classical and quantum worlds, the quantization of energy can only occur in physical system in which particles move randomly.

**A** classical particle without random motion must obey the classical Hamiltonian equation, the particle's energy at any moment is a function of its coordinate q and momentum p (p=$\frac{dq}{dt}$ m). In classical physics, spaces and time are regarded as continuous, so the particle's energy must evolve with time continuously.

However, if we have a physical system in which particles are in random motion, then the energy state of the system should be described by the laws of thermodynamics and statistical physics. The first law of thermodynamics relates the entropy of a system to its internal energy E in differential form as

$$dE = TdS - PdV \qquad (24)$$

Here, the change in internal energy dE is equal to the absolute temperature T times the change in entropy S minus the external pressure P times the change in volume dV. From a statistical viewpoint, the entropy of a system is defined as a measure of randomness or disorder of a system. The more random of the system is, the higher the entropy will be. If the volume of the system remains constant, then the equation reduce to

$$dE = TdS \qquad (25)$$

This equation means that the change of the internal energy of a system is associated with the change of its entropy, integrating both sides of the above equation, we have

$$E = TS \qquad (26)$$

Under the condition of equal probability of outcome, the entropy of the system can be presented as the Boltzmann's constant $k_b$ times the natural logarithm of the number of possible microscopic configuration states (microstates) W, we have

$$S = k_b \ln W \qquad (27)$$

Here, the entropy S depends only upon the state of the system and is independent of path. For a given entropy S of a macro state, there exist a fixed number of the possible microstates W associated with it. Substitution of equation (27) into equation (26) yields

$$E = k_b\, T \ln W \qquad (28)$$

Supposing we have a physical system, in which only two identical particles are in random motion. At low temperature, the system has only two energy levels available for the two particles to occupy. If we put the particles in the way, that at most only one particle can occupy each level, and then the number of the possible microstates W is 2 obviously. Then the ground state $E_0 = k_b T_0$ lu2. With increasing temperature, there will be more vibration energy levels available for the two particles to occupy in the system. Assume that there exist three energy levels at temperature $T_1$, 4 levels at $T_2$, and 5 levels at $T_3$, as well as 6 levels at $T_4$, respectively. Then the corresponding number of the possible microstates for temperature from $T_1$ to $T_4$ can be calculated as following.

Assume that particles are not distinguishable, but energy levels are distinguishable, then putting k particles into n energy levels, with at most one particle in each level, under condition n > k, the number of possible microstates should be $W = \binom{n}{k} = \frac{n!}{k!}$, the n factorial divided by k factorial. It follows from this that the numbers of the possible microstates for temperature from $T_0$ to $T_4$ would be $W(T_0) = 2$, $W(T_1) = 3$, $W(T_2) = 12$, $W(T_3) = 60$, and $W(T_4) = 360$, respectively. Then, the corresponding internal energy of the system becomes $E_0(T_0) = k_b T_0 \ln 2$, $E_1(T_1) = k_b T_1 \ln 3$, $E_2(T_2) = k_b T_2 \ln 12$, $E_3(T_3) = k_b T_3 \ln 60$, and $E_4(T_4) = k_b T_4 \ln 360$, respectively. If we differentiate both sides of equation (28) with respect to temperature T, then

$$\frac{dE}{dT} = K_b \ln W + \frac{K_b T}{W} \frac{dW}{dT} \qquad (29)$$

The first term of the right side of the equation (29) is a constant, and the change of internal energy with temperature is dominated by $\frac{dW}{dT}$, which is the change of microstates with temperature. From the thought experiment, we can see that in a physical system containing a small number of particles, the internal energy of the system will change discontinuously or quantized with increasing temperature.

Now we arrive at a conclusion that the phenomenon of quantization of energy is by no means unique to quantum mechanics, under some special conditions, it may also exist in classical system containing random motion particles, due to the microstates changing discontinuously with temperature.

In quantum mechanics, the wave function describing random motion of particles dominates the energy states of a system. In thermodynamics, when the volume of a system holds constant, its entropy determines the internal energy of the system, which is a measure of the randomness or microstates of random motion particles in the system. It can be seen that the entropy in thermodynamic and the wave function of the Schrödinger equation have one important nature in common; they both are the macro state functions of randomly moving particles, leading to the quantization of energy in classical and quantum system, respectively. Therefore, we conclude that the quantization of energy can only occur in physical system in which particles perform random motion, and that the origin of energy quantization in both classical and quantum system stems from discontinuous changes of the microstates in both systems. For instance, when the microstate of the wave function in an atomic system jumps from an s state to a p or d state, then the corresponding transition occurs in the energy state of the atom.

## IX   Conclusions

The nature of quantum particles is their random motion, which dominates the probabilistic nature of quantum mechanics. Schrödinger's great contribution to modern physics lies in the use of linear differential equation to describe physical system in which subatomic particles move randomly, thereby leading to quantum mechanics.

The Schrödinger equation in the quantum world is equivalent to the Hamilton equation in the Newton world. The two equations are designed to describe deterministically the development of energy of states in course of time. Different from the classical mechanics, the reality described by the wave function of Schrödinger equation is the random motion of particles. Thus, the characteristic of quantum mechanics has a dual nature. On the one hand, quantum mechanics inherits the deterministic nature of classical physics, which is why the predictions of quantum mechanics have been experimentally verified with extremely high accuracy. Quantum mechanics, on the other hand, has inherently probabilistic nature caused by random motion of particles. Keeping this concept in mind, open questions in quantum mechanics, such as wave – particle duality, the quantum superposition principle, double-slit experiments, etc, can all be explained consistently.

Current quantum mechanics is a mixture of matrix mechanics and wave mechanics, both of which are contradictory in principle. Matrix mechanics treats quantum particles as classical particles that satisfy the classical Hamiltonian equation modified by Heisenberg's quantum condition. There exists a fixed relation between position of a particle and its momentum, which belong to measurable physical quantities in matrix mechanics. The measurement of these two physical quantities can certainly determine the energy state of the particle. Both Born's non-commutative relation and Heisenberg uncertainty relation originate from matrix mechanics.

However, the wave mechanics - Schrödinger equation is built on a physical system in which particles move randomly. There is no fixed relation between a particle's positions and its momentum, the coordinate and momentum of a particle are hidden variables in its wave function, which belong to immeasurable physical quantities. Thus, there is no so-called measurement problem such as wave function collapse and many worlds. Therefore, there is no need for non-commutative relation and uncertainty relation in wave mechanics.

In fact, Heisenberg's quantum condition is just the Bohr – Sommerfeld quantum condition in matrix representation. However, the Bohr –Sommerfeld quantum condition does not exist in quantum mechanics. Therefore, we can safely say that, in principle, matrix mechanics should belong to the realm of the old quantum theory. Equations such as non-commuting relation and uncertainty relations derived from matrix mechanics belong to the old quantum theory and are incompatible with wave mechanics.

Based on the discussion above, we conclude that, in principle, non-commuting relation and Heisenberg uncertainty relation should not exist in wave mechanics. In the history of quantum mechanics, it is a fundamental mistake to regard the coordinate and time independent variables of the wave function as the operators of quantum mechanics. They have no physical realities associated with them. Thus, both the non-commuting relations $[\hat{p}, \hat{x}] = -i\hbar$ and $[\hat{E}, \hat{t}] = -i\hbar$ are not real. Without these two non-commuting relations, Heisenberg's uncertainty relations have no mathematical basis for their existence in quantum mechanics.

We propose that the quantization of energy can only occur in the physical system in which particles perform random motion, and that the origin of energy quantization roots in the discontinuous changes of microstates in random-motion system.

In fact, the Schrödinger equation can provide accuracy results for describing electron systems. For other particles, it merely gives approximate results. There is no so-called macroscopic quantum mechanics or universal wave function in nature. Conversely, if subatomic particles stop moving randomly, or move at very high speed, then all subatomic particles, no matter how small they are, must obey the laws of classical physics.

We believe that the unification of the classical world and the quantum world can be achieved on the base of unified physical foundation only when the Heisenberg uncertainty relations are completely abandoned by physical world.

**Acknowledgement**

We would like to thank the authors of the articles cited in this article. The senior authors of this article would like to express particular gratitude to Professor Heribert Wiedemeier, Professor John Schroeder, and Professor Peter Persans for their support during their work at Rensselaer Polytechnic Institute, Troy, NY. USA.